\documentclass[aps,showpacs,showkeys,amsmath,amssymb]{revtex4}

\usepackage{graphicx}% Include figure files
\usepackage{dcolumn}% Align table columns on decimal point
\usepackage{bm}% bold math
\usepackage{mathrsfs}
\usepackage{epsfig}

\newcommand{\N}{N_c}

\begin{document}
%UMD-DOE/ER/40762-508 Nov 16 2011

\title{Total nucleon-nucleon cross sections in large $\N$ QCD}

\author{Thomas D. Cohen}
\email{cohen@physics.umd.edu}
\affiliation{Department of Physics, University of Maryland, College Park, MD 20742-4111, USA}
\author{Boris A. Gelman}
\email{bgelman@citytech.cuny.edu}
\affiliation{Department of Physics, New York City College of Technology, \\
The City University of New York, New York, New York 11201, USA}

%\date{\today}

\begin{abstract}

We use contracted spin-flavor symmetry which emerges in the large $\N$ limit of  QCD to
obtain relations between proton-proton and proton-neutron total cross sections for both polarized and unpolarized scattering.  The formalism used is valid in the semi-classical regime in which the relative momentum of the incident nucleons is much larger than the inverse size of the nucleon, provided that certain technical assumptions are met. The relations should be phenomenologically useful provided that $\N=3$ is sufficiently large so that the large $\N$ results have at least semi-quantitative predictive power. The relations are model-independent in the sense that they depend on properties of large $N_c$ QCD only and not on any particular model-dependent details of the nucleon-nucleon interaction. We compare these model-independent results to experimental data. We find the relation for spin-unpolarized scattering works well empirically.  For the case of polarized scattering, the data is consistent with the relations but the cross sections are too small to make sharp predictions.

\end{abstract}

% PACS, the Physics and Astronomy
                             % Classification Scheme.
%Use showkeys class option if keyword
                              %display desired
\maketitle

\section{\label{sec:intro} Introduction}

Quantum chromodynamics (QCD) is the fundamental theory of the strong
interaction. However, since  standard perturbation theory breaks down
at low momentum transfers,  it is important to develop non-perturbative
techniques to analyze hadronic and nuclear properties. One such  method,
proposed by 't Hooft, is based on a $1/\N$ expansion around the limit
in which the number of colors $\N$ is taken to infinity and the ratio
$g_s/\sqrt{\N}$ is kept constant \cite{tHooft}. The large $\N$ limit and
$1/\N$ expansion has proven to be useful in our understanding of hadrons.

The description of mesons and baryons in the large-$\N$ limit
requires different techniques. A key consequence of large-$\N$
counting rules in the meson sector is that leading order
contributions to the observables come from planar Feynman diagrams \cite{tHooft}.
This allows one to analyze the $\N$ dependence of correlation functions with
quark-antiquark quantum numbers and deduce the $N_c$ scaling of observables.
In particular, large-$\N$ scaling of meson masses and n-meson couplings are
$\N^0$ and $\N^{1-n/2}$, respectively \cite{tHooft}. The latter scaling validates
the OZI rule.

Dominance of planar diagrams is not however enough
to describe the baryon sector of the large-$\N$ QCD. The reason
is that unlike mesons which have quantum numbers of quark-anti-quark
pairs, baryons in the large-$\N$ limit contain $\N$ quarks. As
a result, the baryon observables receive contributions from Feynman
graphs with ever increasing number of quark lines. Thus, a nucleon
is a many-body state and contributions from n-body quark forces
to nucleon observables are of order $\N$. As noted by Witten, this
is precisely the conditions of applicability of mean-field methods \cite{Witten}.
The mean-field equations were only explicitly derived in
the case of non-relativistic heavy quarks in which each quark moves in an average
potential created by $\N-1$ quarks and Hartree mean-field framework is valid.
The solution to these equations were explored only recently\cite{CohenKumarNdousse}.
While  the explicit equations are unknown for the case of light quarks,
Witten argued that large $\N$ scaling of various observables
remains the same.  Witten also noted that baryons at large $\N$
behave analogously to semi-classical solitons. This
solitonic nature of baryons in the large-$\N$
limit will play a crucial role in our treatment of the
nucleon-nucleon scattering observables.

A key feature of the baryon sector in the large-$\N$ limit of QCD is the emergence of a contracted $SU(2N_F)$ spin-flavor symmetry
of the ground state band of baryons \cite{SU2Nf}. In this paper we focus on the implications of  this symmetry on
the strong interaction between two nucleons---a subject of critical importance in nuclear physics.  There has been a certain amount
of study of this problem over the years, much of it focused on the nucleon-nucleon potential \cite{NNpotential}.  While these studies are interesting and the patterns predicted from large $\N$ can be identified in phenomenological potentials, there are a few conceptual issues which cloud these predictions.  The first is simply that the nucleon-nucleon potential is not a true observable but rather is a theorist construct so the ``data'' used in the comparisons is not directly data.  Moreover, in the large $\N$ limit, the $\Delta$ becomes stable and degenerate with the nucleon and this means that the potential one uses in the two-nucleon problem should be the one appropriate for a coupled channel problem including  explicit $\Delta$ baryons, while the phenomenological potentials to which they are compared have the $\Delta$ baryons integrated out.  A final concern is simply the scales in the problem. Note that in the analysis it is implicitly assumed that $\N$ is large enough to justify the approach for nuclear obseravbles for $\N=3$.  However, it seems likely that while the expansion may be useful for typical hadronic observables, the nuclear scales are much smaller for reasons unconnected to large $\N$ and the expansion may not be valid. To see the possible difficulty with treating the physical world as being similar to the large $\N$  world, note that the potential scales as $\N$ (as does the baryon mass) which implies that at large $\N$ the binding energy of nucleons is also of order $\N$.  In practice, however,  the deuteron is barely bound with a binding energy of $2.2 \, MeV$.  In contrast, the nucleon-$\Delta$ mass splitting is neglected in the analysis as a $1/\N$ effect but is about
$300\, MeV$.  It seems problematic to neglect the  nucleon-$\Delta$ mass splitting as small while taking seriously a ``large'' potential yielding a $2 \, MeV$ binding.

In any event, it is interesting to consider whether one can use the contracted $SU(2N_F)$ symmetry implicit at large $\N$  to learn anything  directly about observables associated with the nucleon-nucleon interaction. In this article we consider implications of large-$\N$
scaling rules and of contracted $SU(4)$ symmetry on the nucleon-nucleon scattering observables. At first sight it may seem that this task is hopeless.  As noted  by Witten, the natural large $\N$  description of baryon-baryon scattering is given in terms of  time-dependent mean-field theory (TDMFT) with the velocity as opposed to the momentum held fixed as $\N$ is taken to be large.  Note that this implies that the scattering is essentially semi-classical in nature at large $\N$---the relative momenta is much larger than the inverse size of the interaction region.   As we will discuss below, the restriction to the semi-classical scattering regime will ultimately complicate the quest for testable predictions.    However, TDMFT describes averages over processes and thus there is apparently no way to compute S-matrix elements directly in the mean-field framework \cite{Griffin}. The contracted $SU(2N_F)$ symmetry relates observables for baryons with different spins and isospins. Since in TDMFT one has no access to the S-matrix elements any mean-field treatment will have no sensitivity to the spin and isospin of the final baryons after the scattering, and hence no way to impose the symmetry properties on the final state baryons.

However, despite the above limitations, it {\it is} possible to use the emergent symmetry to make testable predictions associated with nucleon-nucleon scattering \cite{NNobservables,CD}. These predictions become exact as $\N \rightarrow \infty$ with the relative velocity of the initial baryons held fixed.  There  are certain inclusive observables which are in principle calculable from TDMFT and which correspond to weighted averages over the sum of the square of certain S-matrix elements. The most basic observable we can compute in  TDMFT is the net collective flow of a conserved quantity such as energy density or baryon density \cite{NNobservables}. These observables sum over many physical final states. While one cannot exploit the contracted $SU(2N_F)$ symmetry on the final state, one can for the initial state. As a result, one can relate the flow observables for different spin and isospin configurations for the initial state in a model-independent way \cite{NNobservables}.

It would be of interest to directly test these predictions. Data exists for the various initial spin and isospin channels. However, the data is not conventionally presented in the form of flow observables making it cumbersome to do such a test.  An alternative approach would be to use kinematics to simplify the analysis \cite{CD}. One could ask what happens for small velocities. At sufficiently small velocities, there is not enough kinetic energy for any inelastic processes to occur. The inelastic threshold at large $\N$ given in terms of relative velocity is
\begin{equation}
v_t = \sqrt{ \frac{4 m_\pi}{M_N}} \sim \sqrt{\frac{1}{N_c}} \; .
\end{equation}
For $v<v_t$ the only allowed processes are elastic and plenty of elastic scattering data are available which in principle can be used to  test the predictions. Of course there is a restriction as to how small one could go while remaining in the regime of validity of TDMFT.  Witten noted long ago that a smooth large $\N$ limit holds for $v$ held fixed as $\N \rightarrow \infty$ and it has been conventional to regard such scaling as necessary \cite{Witten}. With such a scaling rule one finds that as $\N \rightarrow \infty$, $v_t < v$ for any nonzero velocity and thus it looks as though the elastic region is excluded at large $\N$.  However, this is not really the case.  One can ask what  happens if the velocity at large $\N$ approaches zero, but does so in a way that keeps the system in the regime of validity of TDMFT which is ultimately the semi-classical regime.  The condition for being in this regime is that the momentum is much larger than the range of the interaction.  In terms of the velocity this amounts to the condition that relative velocity is parametrically larger than a quantity of order $1/\N$.  Thus at large $\N$ there exists a regime where $v$ is large enough to be in the regime of validity of TDMFT while still being below the elastic threshold.

Thus to the extent that $\N=3$ is large enough, one can test the model independent relations by using elastic scattering data just below the threshold for pion production.  At a phenomenological level these relations fail badly indicating that $\N=3$ is not large enough \cite{CD}.  This is not surprising for two reasons.  The first is that in this regime it is not $\N^{-1}$ which acts as the expansion parameter but $\N^{-1/2}$; for $N_c$ as small as 3 this is a rather dubious expansion even for the purpose of qualitative studies. The problem is compounded by the fact that the pion is a pseudo-Goldstone boson and thus has a mass which is anomalously light on the scale of QCD.  This means that the elastic threshold is anomalously light for reasons which have nothing to do with large $\N$ which restricts the domain of validity.  Given these two facts it is understandable why the prediction for a truly large $\N$ for these elastic scattering observables are not relevant at $\N=3$

This paper seeks a method to test model-independent results in a regime in which the natural expansion parameter is $1/\N$, which has no unnaturally light scales and for which there exists a set of analyzed data. Total nucleon-nucleon scattering cross sections would seem ideal for this purpose except for an apparently fatal flaw:  the semi-classical approach on which the analysis is based, is known to fail for scattering at nearly forward angles \cite{L&L}. Since the total cross section includes these forward angles, it would seem that it is unsuitable for a mean-field treatment. Indeed, the classical cross sections diverge due to the contributions of nearly forward scattering and thus one cannot compare the classically computed infinite total cross sections with the finite cross sections obtained from an experiment. The reason the classical total cross section are divergent is quite simple and can be easily illustrated for non-relativistic point-particle scattering. In classical dynamics there is a contribution to the scattering from {\it any} impact parameter, since no matter how large it is, it will lead to {\it some} (albeit very small) deflection of the particle and hence will contribute to nearly forward scattering.

The purpose of this paper is to show that despite the fact that the semi-classical approach implicit in the large $\N$ analysis cannot be used to directly compute the total cross section---even in principle---that it is never-the-less entirely suitable for deducing the spin and isospin dependence of the total cross section, provided that certain technical assumptions hold. In addition, we will analyze the total cross section data to test the large $\N$ predictions arising from the spin-flavor symmetry. The critical observation underlying this analysis is that the inclusive differential cross section at large $\N$ {\it is} computable in TDMFT {\it except} for very forward scattering. Now the total cross section is obtained by integrating the inclusive differential cross section over all angles.  Suppose that the  integral for the exact quantum mechanical theory at large but finite $\N$ is dominated by a region which excludes the very forward angles where the semi-classical analysis breaks down. This is very plausible, since at large $\N$, the angular region where the semi-classical region breaks down becomes very small. If this is true, then the spin-flavor relations controlling the inclusive differential cross section go over to the total cross section up to small corrections which vanish at large $\N$.

With this insight we obtain predictions relating proton-proton with proton-neutron scattering for the unpolarized, longitudinally and transversely polarized total cross section which should be valid at large $\N$ at sufficiently high momentum so that the semi-classical analysis holds.  The prediction works quite well for the unpolarized cross section where the data show that the proton-proton and proton-neutron cross sections are very similar. For the longitudinally and transversely polarized cases, the proton-proton cross sections are quite small on the natural scale of the problem (the unpolarized cross section). This means that the coefficient of the leading-order term in the $1/\N$ expansion is unnaturally small and we can make no sharp predictions. However, we {\it can} make a qualitative prediction. Since the same leading order term controls proton-neutron scattering, a small proton-proton polarized total cross section implies a small proton-neutron  polarized total cross section---a fact borne out by the experiment.

This paper is organized as follows. In the next section we briefly review the application of TDMFT which is used to deduce spin-flavor relations for the inclusive differential cross section. In Sec.~\ref{sec:total} we present a detailed argument for why semi-classical treatment and the spin-flavor symmetry applies to the total cross section. Finally, in Sec.~\ref{sec:experiment}, we discuss the large-$\N$ predictions in light of the experimental data.

\section{\label{sec:semiclassical} Time-dependent mean field framework}

In this section we review the analysis of Ref.~\cite{NNobservables} on the TDMFT framework for description of the spin-flavor dependence of the inclusive differential cross section. As is the case of a single baryon in the large-$\N$ limit, the dynamics underlying baryon-baryon interaction is that of many quarks and gluons interacting among themselves. An appropriate description in this case is
the time-dependent mean-field theory at a fixed baryon velocity \cite{Witten}. In this framework each quark and gluon moves
in an average time-dependent field created by all other particles. This mean-field treatment is essentially classical in nature.

One important fact is that there are classically flat directions in the dynamics \cite{ZeroModes,ZeroModesTwo}.  These are associated with collective degrees of freedom.  The dynamics of these are slow compared to the typical degrees of freedom in the problem (typically down by $1/N_c$ and  this scale separation allows one to isolate the dynamics of the collective degrees of freedom from the full problem.   This allows one to treat the intrinsic degrees of freedom classically while requantizing the collective degrees of freedom.  This is critical since the mean-field treatment always breaks symmetries and these breakings always lead to collective degrees of freedom.  The requantization of  these restores the symmetries and allows one to compute observables associated with states with good quantum numbers.

The Skyrme model \cite{Skyrme,SkyrmeReview}, while unrealistic in detail, is a good paradigm for how this works. It has long been known \cite{ANW} that there are relations between observables in the Skyrme model which follow entirely from the collective degrees of freedom and are independent of all details of the model \cite{BaryonsModelIndependent}. It was subsequently shown that these model-independent relations follow from a contracted $SU(2N_f)$ symmetry which emerges for baryons in the large $N_c$ limit of QCD \cite{SU2Nf}.

At present, the  large-$\N$ TDMFT equations for baryon-baryon scattering in QCD  are unknown. However, as
was discussed in Ref.~\cite{NNobservables}, one can exploit the spin-flavor structure of the collective degrees of freedom to obtain model-independent relations between some observables. Again a simple way to illustrate this is through the Skyrme model.  While there have been numerical simulations of skyrmion-skyrmion scattering in TDMFT \cite{SkyrmionSkyrmion}, these are not of direct interest here as they depend on the model details. The focus here are on those features which are independent of the model details and which are a direct consequence of the large $\N$ structure built into the model.

To obtain the model-independent relations between relevant nucleon-nucleon scattering observables the latter should be, at least in principle,
calculable in TDMFT. Thus, an important question is what class of observables can be defined in TDMFT and what do they correspond to in the full quantum theory? To apply TDMFT for skyrmion-skyrmion scattering one needs to start with initial conditions corresponding to two skyrmions moving with a velocity $v/2$ towards each other separated by an impact parameter $b$. Skyrmions, however, are not nucleons---they are hedgehogs corresponding to classical field configurations which, up to collective space and isospace rotations given by an $SU(2)$ matrix-valued field, describe pion degrees of freedom $U_{h}(\vec{r})= {\rm exp}\left({\rm i} \vec{\tau}\hat{n}F(r)\right)$. These configurations correspond to superpositions of nucleon and $\Delta$ states (as well as other baryons from a ground state band with spin-isospin $I=J=5/2, ...\, \N/2$ in the large $\N$ limit). Since these states become degenerate at large $\N$, the space and isospace rotation of the hedgehog is slow. One can therefore associate an adiabatic collective degrees of freedom $A(t) \in SU(2)$ describing slowly rotating hedgehog configurations $U(r,t) \rightarrow A\dagger(t)  U_{h}(r) A(t)$. The key to the rest of this analysis is that the two initial hedgehogs can have different values of the collective degrees, {\it i.e.}, have different orientation in space and isospace. Thus, in a fully quantal setting the initial conditions correspond to spatial wave packets of hedgehog superpositions with some initial orientation in space moving towards each with an impact parameter $b$.
An ability to construct appropriate initial states are not enough however to extract meaningful quantal information from TDMFT calculations. Since at large-$\N$ the nucleon-nucleon scattering at fixed velocity is in a semi-classical regime, mean-field framework corresponds to the semi-classical treatment. Moreover the nature of TDMFT is such that one can only obtain an information associated with average flows of quantities such as energy or baryon number, but not particular s-matrix elements.

To determine the spin-flavor structure of various total cross sections in TDMFT one needs to construct operators which depend on collective degrees of freedom and which after quantization will correspond to appropriate inclusive observables. Using semi-classical quantization techniques one can extract information about nucleons with particular spin orientations from calculations based on rotated hedgehogs.
For a generic scattering observable this was done in Ref.~\cite{NNobservables} where  a corresponding operator was obtained from
a conserved current. In the Skyrme model the conserved current is topological in nature and is associated with a baryon number. In addition to collective coordinates the conserved current is a function of other ({\it intrinsic}) degrees of freedom which determine its time-evolution.
The value of the current depends on the initial conditions which as discussed above correspond to two well-separated hedgehogs moving toward each other. In addition to two matrices $A_{1,2}$ determining orientation of hedgehogs, the separation between them, their impact parameter and their relative velocity serve to specify the collective degrees of freedom. As a result, at the classical level we have a function $B_{\mu}\left(\vec{r}, t; \vec{b}, \hat{n}, v, A_1, A_2\right)$ where the initial separation between hedgehogs is suppressed since it is irrelevant in the following analysis.

Since scattering describes the long time behavior of the system an appropriate observable at long times corresponds to the outward flow of the baryon number. This outward net flow of a baryon number at a fixed solid angle $\Omega$ is given by
\begin{equation}
\frac{dN^{B}(v, b, A_1,A_2; \theta,\phi)}{d\Omega} = \lim_{R \rightarrow \infty}
{R^2}\int^{\infty}_{0} dt
\hat{r} (\theta,\phi) \cdot
\vec{B}\left(t, R\hat{r}(\Omega); \vec{b}, \hat{n}, v, A_1, A_2\right) \,,
\label{DifferentialBaryonNumber}
\end{equation}
where polar angles $\theta$ and $\phi$ specify the direction of the outgoing current and by construction the time $t=0$ corresponds to the time at which the total baryon density  has the smallest RMS radius ({\it i.e.}, when the two baryons are the closest). The restriction to positive times enforces the condition that we are tracking the outgoing motion of the baryons. The observable defined in Eq.~(\ref{DifferentialBaryonNumber}) is designed to track the outgoing direction of the baryons and it gives the net baryon number flow outward through a given differential element of a solid angle $d \Omega$. It is normalized so that
$\int d \Omega\left( dN^{B}/d\Omega \right) =2$ since the net number of outgoing baryons is 2.

However, $dN^{B}/d\Omega $ is not a cross section. It depends on the impact parameter $b$ as well as on the collective spin-flavor variables $A_1$ and $A_2$. Never-the-less, it is trivial to convert it into a certain type of inclusive differential cross section by integrating over impact parameter space:
\begin{equation}
\frac{d \sigma_{\rm inc}  (v,A_1,A_2;\theta,\phi)} {d\Omega}=
\int_0^\infty  db \, (2 \pi b)  \,
\frac{dN^{B} (v, b, A_1,A_2;\theta,\phi)}{d\Omega}
\label{hhscat} \; .
\end{equation}
Physically, $d \sigma_{\rm inc}/d\Omega$ corresponds to the cross section for one baryon to emerge in a cone of angular size $d \Omega$ about a specific direction integrating over all other variables---the energy of the baryon, the number and kinematics of outgoing mesons, the energy of and direction of the other baryon, as well as isospin and other baryon quantum numbers.

It is important to stress that the inclusive differential cross section defined in Eq.~(\ref{hhscat}) describes not the nucleon-nucleon but rather a hedgehog-hedgehog scattering since it depends on the collective variables $A_1$ and $A_2$. The hedgehog-hedgehog scattering is a well-posed problem in the  large $\N$ limit where the various baryons composing the two hedgehogs become degenerate and two hedgehogs are sensible as an asymptotic state. To turn this cross section into a corresponding nucleon-nucleon one, all that needs to be done is to evaluate $A_1$ and $A_2$ from parameters specifying the quantum collective variables and then calculate the expectation value
of the cross section in Eq.~(\ref{hhscat}) in the quantum state appropriate for particular nucleon spin-isospin quantum numbers. Since in
the initial state the hedgehogs are well separated they can be quantized independently. To evaluate the above expectation value
one needs a nucleon-wave function in the space of collective rotations parameterized by the parameters of $A$-matrices. As is well known \cite{ANW}, these wave functions are given in terms of Wigner $D$-matrices, namely $D^{1/2}_{m_, m^I}(A)$. Accordingly,
the inelastic differential cross section for the two-baryon initial states with spin and isospin projections
$J_{z1}=m_1, I_{z1}=m^{I}_1, J_{z2}=m_2, I_{z2}=m^{I}_2$ can now be found by integrating over the $SU(2)$ measure:
\begin{equation}
\frac{d\sigma^{(m_1, m^I_1, m_2 , m^I_2)} (v, \theta,\phi)}{d\Omega} = \\
\int dA_1 dA_2 |D_{m_1,m^I_1}^{1/2}(A_1)|^2 |D_{m_2,m^I_2}^{1/2}(A_2)|^2
\frac{d \sigma_{\rm inc}  (v,A_1,A_2;\theta,\phi)} {d\Omega} \,.
\label{sigmaInclusive}
\end{equation}

Integration over impact parameter space in Eq.~(\ref{hhscat}) and over the $SU(2)$ measure in Eq.~(\ref{sigmaInclusive}) can be done using time reversal and parity invariance and exploiting the fact that $\left(D^{J}_{m, n}\right)^{*}= (-1)^{m-n} D^{J}_{-m,-n}$. This yields the following structure:
\begin{equation}
\begin{split}
&\frac{d\sigma^{(m_1,m^I_1, m_2, m^I_2)} (v,\theta,\phi)}{d\Omega}= \\
&\left \langle   m_1,m^I_1, m_2 m^I_2  \left| a_0(v, \theta,\phi) + b_I(v, \theta,\phi) \, \left(\vec{\sigma}_1\cdot\vec{\sigma}_2\right)\,
\left(\vec{\tau}_1\cdot\vec{\tau}_2\right) +
c_I(v, \theta,\phi)\, \left(\vec{\sigma}_1\cdot\vec{n}\right)
\left(\vec{\sigma}_2\cdot\vec{n}\right)\,
\left(\vec{\tau}_1\cdot\vec{\tau}_2\right) \right |  m_1,m^I_1, m_2 m^I_2 \right \rangle \; ,
\end{split}
\label{DSigma}
\end{equation}
where the $\sigma$ and $\tau$ matrices are the standard Pauli matrices which act on two-baryon states of the form
$|m_1,m^I_1, m_2 m^I_2 \rangle $, and the functions $a_0$, $b_I$ and $c_I$ encode the leading order behavior at large $N_c$.

Note that the form of Eq.~(\ref{DSigma}) is completely determined by large-$\N$ consideration and the spin-flavor symmetry implicit therein.  It is independent of the details of the Skyrme model and thus one expects that it is a consequence of large-$\N$ QCD itself.  The detailed form of the functions $a_0$, $b_I$ and $c_I$ are of course model dependent and one cannot deduce them from general considerations alone. However, the form of Eq.~(\ref{DSigma}) contains important information.  It is {\it not} the most general form one can write consistent with parity and time reversal.  For example, there is no term of the form $a_I \left(\vec{\tau}_1\cdot\vec{\tau}_2\right)$. This means that one can make concrete predictions based on the form.  Of course, in doing so one needs to remember two key things.  Firstly, that this is only the leading term in the $1/N_c$ expansion and thus predictions based on the form are not exact due to $1/N_c$ corrections.  Secondly, the result only applies in the semiclassical regime.

\section{\label{sec:total} Applicability to total cross sections}

In the previous section we derived an expression for the inclusive differential cross section Eq.~(\ref{DSigma}), which is valid in the semiclassical regime at large $\N$.  The purpose of this section is to argue that an analogous result holds for total cross sections, namely
\begin{equation}
\begin{split}
&\sigma^{(m_1,m^I_1, m_2 , m^I_2)} (v) = \\
&\left \langle   m_1,m^I_1, m_2 , m^I_2  \left| A_0(v) + B_I(v) \, \left(\vec{\sigma}_1\cdot\vec{\sigma}_2\right)\,
\left(\vec{\tau}_1\cdot\vec{\tau}_2\right) +
C_I(v)\, \left(\vec{\sigma}_1\cdot\vec{n}\right)
\left(\vec{\sigma}_2\cdot\vec{n}\right)\,
\left(\vec{\tau}_1\cdot\vec{\tau}_2\right) \right |  m_1,m^I_1, m_2, m^I_2 \right \rangle \; .
\end{split}
\label{Sigma}
\end{equation}
Note that the total cross section is related to the inclusive differential cross section through
\begin{equation}
\sigma^{(m_1,m^I_1, m_2 , m^I_2)} (v) =\frac{1}{2} \int d\Omega \, \frac{d\sigma^{(m_1,m^I_1, m_2 , m^I_2)}
(v, \theta,\phi)}{d\Omega} \; ,
\label{tot1}
\end{equation}
where the angular integral is over $4 \pi$ and factor of $1/2$ comes from the normalization of $dN^{B}/d\Omega$ in
Eq.~(\ref{DifferentialBaryonNumber}) and accounts for the fact the baryon number of the system is 2. Integrating both sides of Eq.~(\ref{DSigma}) over angles and exploiting Eq.~(\ref{tot1}) immediately yields Eq.~(\ref{Sigma})
with
\begin{equation}
\begin{split}
&A_0 (v) =\int d\Omega \,  a_0(v;\theta,\phi)  \\
&B_I(v) =\int d\Omega \,  b_I(v;\theta,\phi)  \\
& C_I(v) =\int d\Omega \,  c_I(v;\theta) \,.
\end{split}
\end{equation}
Unfortunately, there is a problem with this. Equation (\ref{DSigma}) only holds in the semiclassical limit and that excludes very forward angles while the integration in  Eq.~(\ref{tot1}) is over {\it all} angles including forward and backward ones. Note forward scattering contributes at both forward and backward angles since $d\sigma^{(m_1,m^I_1, m_2 , m^I_2)}/d\Omega$ includes both of the outgoing baryons.

The problem of forward scattering however need not invalidate Eq.~(\ref{Sigma}) (as a result valid at leading order in $1/\N$). Suppose that for the full quantum problem at large but finite $\N$, the angular integral in Eq.~(\ref{tot1}) is dominated by angles which {\it are} valid in the semiclassical regime. If this were to happen then one would expect that Eq.~(\ref{Sigma})  would hold up to small corrections.  It is important to make this statement somewhat more precise.  To to do so we introduce the following quantity:
\begin{equation}
\sigma^{\theta_0} =   \pi \, \int_{\theta_0}^{\pi-\theta_0} \, \,\frac{d\sigma (\theta,\phi)}{d\Omega}  \, \sin(\theta) d \theta
\end{equation}
which corresponds to the total cross section {\it excluding} scattering where an outgoing nucleon passes through a cone of angular width $\theta_0$ about the scattering axis. It is easy to see that Eq. (\ref{Sigma}) holds at large $\N$ and some fixed $v$, provided that there exists some function $\theta_0 (\N,v)$ that satisfies two conditions:
\begin{enumerate}
 \item  Scattering with nucleons emerging  with $\theta_0(N_c,v) < \theta <\pi -\theta_0(N_c,v)$ is sufficiently semiclassical that Eq.~(\ref{DSigma}) is accurate (in the sense that corrections to Eq. (\ref{DSigma}) go to zero as $N_c \rightarrow \infty$  at fixed $v$ for all angles in this window).
\item  The full quantum cross sections satisfy
\begin{equation}
\lim_{N_c \rightarrow \infty}  \frac{ \sigma^{\theta_0(N_c,v)} }{\sigma} \rightarrow 1 \; .
\end{equation}
\end{enumerate}

The issue is whether a function $\theta_0 (N_c,v)$ satisfying these two conditions exists.  The following argument suggests that it is highly plausible that it does. To begin, note that in the semiclassical regime,  the outgoing angle at which the baryon emerges is determined by the impact parameter with forward angles associated with large impact parameters.  Thus, condition 1 translates into a question of how large can the impact parameter be at given $v$ and $N_c$ while still being in the semiclassical limit. Now for {\it any} fixed value of $v$ and any fixed $b$, Witten's \cite{Witten} reasoning implies there must be some value of $N_c$ for which  the scattering is semiclassical. Because the interaction strength falls off like a Yukawa potential at large distances, as $b$ increases beyond the characteristic range of the interaction, the value of $N_c$ needed to be in the semiclassical regime will grow very rapidly with $b$. Nevertheless, one can always go to sufficiently large $N_c$ so that any given $b$ (and hence any fixed scattering angle) is accurately described semiclassically.  This in turn implies that there must exist a function $\theta_0(N_c,v)$ which satisfies condition 1 and also has the property that $\lim_{N_c \rightarrow \infty} \theta_0(N_c,v)\rightarrow 0$.  That is, as $N_c$ goes to infinity, the angular region of validity of the semiclassical region approaches $4 \pi$. This in turn implies that condition 2 is met unless the scattering becomes so forward-peaked at large  $N_c$ that the dominant scattering occurs in the infinitesimally small region of angles less than $\theta_0(N_c,v)$. This would require an exceptionally forward-peaked cross section dominated by elastic scattering. In this paper, we will assume that such extreme forward peaking does not occur at large $N_c$.  We  base this conclusion on phenomenology since in the regime we study elastic scattering as a small fraction of the total. The theoretical question of whether this assumption is correct in the formal large $N_c$ limit is interesting and will be pursued in future work. We note here however, that for the assumption to be wrong, the cross section would have to be extremely forward peaked to a degree that seems {\it a priori} implausible.

\section{\label{sec:experiment} Comparison with Experimental Data}

In this section we compare the leading order form of the total nucleon-nucleon
crosse section at large $\N$, Eq.~(\ref{Sigma}), with data from  nucleon-nucleon scattering
for center-of-mass momenta of a few $GeV$. This is the energy scale  for which the system is
in the semi-classical regime for all but the most forward angles.

The data for the total cross sections is usually quoted for particular isospin channels.
Using isospin projection operators
$\left(1-\vec{\tau}_1\cdot\vec{\tau}_2\right)/4$ and
$\left(3-\vec{\tau}_1\cdot\vec{\tau}_2\right)/4$ one can extract from
Eq.~(\ref{Sigma}) the cross sections for isosinglet and isotriplet channels,
\begin{eqnarray}
\sigma^{(I=0)} & = &
\left(A_0 - 3 B_I \, \left(\vec{\sigma}_1\cdot\vec{\sigma}_2\right)-
3 C_I\, \left(\vec{\sigma}_1\cdot\vec{n}\right)
\left(\vec{\sigma}_2\cdot\vec{n}\right)\right)
\nonumber \\
\sigma^{(I=1)} & = &
\left(A_0 + B_I \, \left(\vec{\sigma}_1\cdot\vec{\sigma}_2\right) +
C_I\, \left(\vec{\sigma}_1\cdot\vec{n}\right)
\left(\vec{\sigma}_2\cdot\vec{n}\right)\right) \, .
\label{sigmaI}
\end{eqnarray}

Since reactions $p\,p \to X$, $n\,n \to X$ receive contribution only from
isotriplet channel $\sigma^{pp} = \sigma^{nn} = \sigma^{I=1}$, while the
reaction $n\,p \to X$ receives equal contributions from both channels
$\sigma^{np} =  \frac{1}{2}\left(\sigma^{I=1} + \sigma^{I=0}\right)$,
it follows from Eq.~(\ref{sigmaI}) that at leading order in $1/\N$ expansion,
\begin{eqnarray}
\sigma^{(pp)} & = & \sigma^{(nn)} =
A_0 + B_I \, \left(\vec{\sigma}_1\cdot\vec{\sigma}_2\right) +
C_I\, \left(\vec{\sigma}_1\cdot\vec{n}\right)
\left(\vec{\sigma}_2\cdot\vec{n}\right) \nonumber
\\
\sigma^{(np)} &=&
A_0 - B_I \, \left(\vec{\sigma}_1\cdot\vec{\sigma}_2\right)-
C_I\, \left(\vec{\sigma}_1\cdot\vec{n}\right)
\left(\vec{\sigma}_2\cdot\vec{n}\right) \,.
\label{channelSigmas}
\end{eqnarray}

Data exists for both spin-averaged and polarized nucleon-nucleon scattering cross sections \cite{NNexperimental}.
A general form of a total cross section for two spin-1/2 particles as a function of the initial particle polarizations
is
\begin{equation}
\sigma =\sigma_0 + \sigma _1 \left(\vec{P}_B\cdot\vec{P}_T\right) +
\sigma_2 \left(\vec{P}_B\cdot\hat{n}\right)
\left(\vec{P}_T\cdot\hat{n}\right) \,,
\label{polSigmaGeneral}
\end{equation}
where $\vec{P}_B$ and $\vec{P}_T$ are polarizations of a beam
and target particles, respectively, $\hat{n}$;
$\vec{P}_B\cdot\vec{P}_T =  < \vec{\sigma}_1\cdot\vec{\sigma}_2 >$ and
$\left(\vec{P}_B\cdot\vec{n}\right) \left(\vec{P}_T\cdot\vec{n}\right)
= < \left(\vec{\sigma}_1\cdot\vec{n}\right)
\left(\vec{\sigma}_2\cdot\vec{n}\right) >$. In Eq.~(\ref{polSigmaGeneral}),
$\sigma_0$ is the spin-averaged total cross section.

It follows from Eqs.~(\ref{channelSigmas})
and (\ref{polSigmaGeneral}) that the spin-averaged total cross section
at leading order in $1/\N$ for all three reactions is the same:
\begin{equation}
\sigma^{(pp)}_{0} = \sigma^{(nn)}_{0} = \sigma^{(np)}_{0} \left(1+{\cal{O}}(1/\N)\right)\,.
\label{predictionSigma0}
\end{equation}
Note, that while the first equality is due to isospin invariance, the second equality
is a prediction of large-$\N$ QCD.

\begin{figure}[htb]
\begin{center}
\epsfig{figure=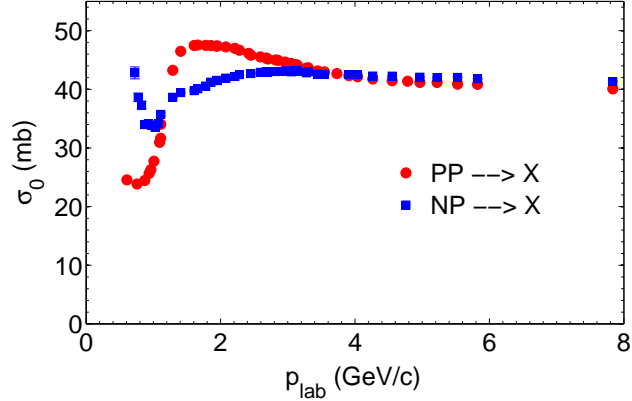,height=55mm}
\end{center}
\vspace{-6mm}
\caption{(Color online) Spin-averaged proton-proton and neutron-proton total cross section as a function
of beam momentum (Bugg et al., 1996).}
\label{SpinIndependentFig}
\end{figure}

The above large-$\N$ result is well satisfied by data as shown in Fig.~\ref{SpinIndependentFig}.

One can also obtain large-$\N$ predictions for polarized cross sections.
In polarized scattering experiments the beam and target nucleons
can have either transverse or longitudinal polarization relative to the
the incident beam direction $\hat{n}$. In addition, the nucleons can be
polarized in the same or opposite relative direction.

\begin{figure}[htb]
\begin{center}
\epsfig{figure=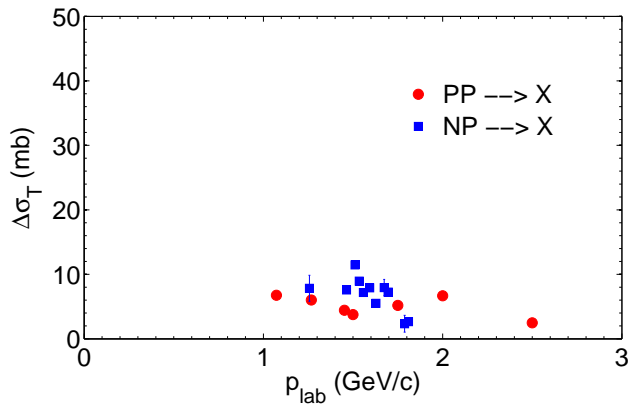,height=55mm}
\end{center}
\vspace{-6mm}
\caption{(Color online) Spin-dependent neutron-proton (Fonteine et al., 1991) and proton-proton
(Ditzler et al., 1983; Lesikar, J. D. 1981) total cross section differences as a function
of beam momentum. $\Delta\sigma_T$ is defined in Eq.~(\ref{DeltaSigmaTransverse}).}
\label{DeltaSigmaTFig}
\end{figure}

It is customary to combine two cross sections for transversely polarized nucleons,
$\sigma(\uparrow\uparrow)$ and $\sigma(\uparrow\downarrow)$, into an observable
referred to as {\it delta sigma transverse} defined as
\begin{equation}
\Delta\sigma_T = -\left(\sigma(\uparrow\uparrow)- \sigma(\uparrow\downarrow)\right)
=-2 \sigma_1\,,
\label{DeltaSigmaTransverse}
\end{equation}
where the last equality follows from Eq.~(\ref{polSigmaGeneral}).
Using Eq.~(\ref{channelSigmas}) we obtain at leading order in $1/\N$,
$\Delta\sigma^{(pp)}_{T} = \Delta\sigma^{(nn)}_{T} = - 2 B_I$ and
$\Delta\sigma^{(np)}_{T} = 2 B_I$. Thus, up to $1/\N$ corrections
we have the following prediction,
\begin{eqnarray}
\Delta\sigma^{(pp)}_{T} = \Delta\sigma^{(nn)}_{T} = - \Delta\sigma^{(np)}_{T}
\left(1+{\cal{O}}(1/\N)\right)\,.
\label{predictionDeltaSigmaT}
\end{eqnarray}
Data for this observable is shown in Fig.~\ref{DeltaSigmaTFig}.

At first sight data appears to violate the relation in Eq.~(\ref{predictionDeltaSigmaT}) badly. However, this is misleading. Recall that this prediction, is only valid to leading order at large $N_c$. If it happens that the leading order coefficients are anomalously  small for reasons not associated with $N_c$, then one does not  expect the leading terms to dominate at $N_c=3$.  In the present case, the leading order coefficients {\it are} small. Note that the the characteristic size of cross sections in the problem are those of the total cross section,
Fig.~\ref{SpinIndependentFig}, and one sees that $\Delta\sigma^{(pp)}_{T}$ is smaller than $\sigma^{(pp)}$ by a large factor.  This means that the system is likely to be outside of the range of validity of the $1/N_c$ expansion for this observable, and one does not expect the relation to hold quantitatively. As a result, it does not provide a sharp quantitative test of the $ 1/N_c$  expansion.  However, there is a {\it qualitative} prediction that we can test. In particular if  $\Delta\sigma^{(pp)}_{T}$ is much less than the unpolarized cross section then so is $\Delta\sigma^{(pn)}_{T}$ since both follow from the leading order term in the $1/N_c$ expansion for $B_I$ which is small.  This qualitative prediction does indeed hold as seen in Fig.~\ref{DeltaSigmaTFig}.

\begin{figure}[htb]
\begin{center}
\epsfig{figure=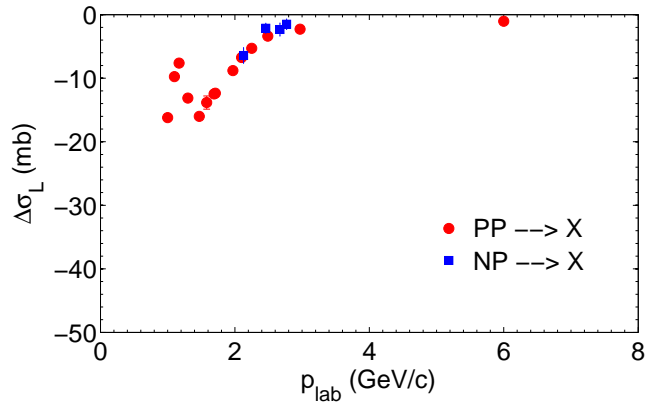,height=55mm}
\end{center}
\vspace{-6mm}
\caption{(Color online) Spin-dependent proton-proton (Auer et al., 1978), and neutron-proton
(Sharov et al., 2008)  total cross section differences as a function
of beam momentum. $\Delta\sigma_L$ is defined in Eq.~(\ref{DeltaSigmaLongitudinal}). }
\label{DeltaSigmaLFig}
\end{figure}

For longitudinal polarization two cross sections $\sigma(\rightrightarrows)$ and $\sigma(\rightleftarrows)$ are
combined to give
\begin{equation}
\Delta\sigma_L = -\left(\sigma(\rightrightarrows)- \sigma(\rightleftarrows)\right)
=-2 \left(\sigma_1 + \sigma_2\right)\,,
\label{DeltaSigmaLongitudinal}
\end{equation}
which is referred to as {\it delta sigma longitudinal}.
Using Eq.~(\ref{channelSigmas}) we obtain at leading order in $1/\N$,
$\Delta\sigma^{(pp)}_{L} =\Delta\sigma^{(nn)}_{L} =- 2 \left(B_I + C_I\right)$
and $\Delta\sigma^{(np)}_{L} = 2 \left(B_I + C_I\right)$. Thus,
at leading order in $1/\N$ expansion,
\begin{equation}
\Delta\sigma^{(pp)}_{L} = \Delta\sigma^{(nn)}_{L}  =  - \Delta\sigma^{(np)}_{L}
\left(1+{\cal{O}}(1/\N)\right) \, .
\label{predictionDeltaSigmaL}
\end{equation}
Data for the above observable is shown in Fig.~\ref{DeltaSigmaLFig}.
Again, in the region where data exists the cross sections are too small for the relations to be expected to hold for $N_c=3$. However, there is a qualitative prediction that if $ \Delta\sigma^{(pp)}_{L}$ is small then so is $ \Delta\sigma^{(pn)}_{L}$. This qualitative prediction holds, as is shown in
Fig.~\ref{DeltaSigmaLFig}.

In summary, we have argued that the spin-flavor symmetry which emerges in the large $N_c$ limit of QCD allows for predictions for total cross sections at sufficiently large initial momenta. The prediction for the spin-averaged cross section works well. The polarized cross sections appear to be too small to be in the regime of validity of the $1/N_c$ expansion with $N_c=3$.  However, this fact itself allows for a qualitative prediction which does hold.

\begin{acknowledgments}
BAG would like to gratefully acknowledge the support of the Professional Staff Congress-City University of New York
Research Award Program through the grant PSCREG-41-540. BAG and TDC would like to gratefully acknowledge a support of
the Center for Theoretical Physics, New York City College of Technology, The City University of New York where this work began.  TDC thanks the U.S. Department of Energy under grant DE-FG0293ER-40762 for its support.
%BG would like to thank R. Ya. Kezerashvili for helpful discussions and encouragement.

\end{acknowledgments}

\end{document}